# Loss Minimization through the Allocation of DGs Considering the Stochastic Nature of Units

Ali Parsa Sirat
Aliparsasirat2016@gmail.com

*Abstract*—Smart grid as the cleaner alternative to the legacy power system can improve technical, economical, and environmental aspects of the system up to a considerable degree. In smart grids, Distributed Generation (DG) units; which play an important role, should be optimally allocated. In this paper, DG placement is conducted with the goal of loss minimization of the grid considering the technical limitations associated with the voltage profile of the buses as well as the stochastic nature of the DGs. In this paper, three different kinds of DGs are included which are wind turbines, solar panels, and biomass generators. The results related to the case study which is IEEE standard 33 bus system reveals that the costs can be dramatically decreased.

## I. INTRODUCTION

Nowadays, electrical engineers are facing new challenges for both planning and operation of electric distribution grids [1] including:

1- The more the penetration of the renewable energy resources is, the more balancing supply and demand is challenging [2]- [5]. That is because of stochastic nature of certain renewable energy resources.
2- Especially in recent years, electric vehicles are clean alternatives to conventional internal combustion engine vehicles [6] and hydraulic actuator-based ones (due to the oil leakage) [7], [8]. However, the stochastic nature of driver behaviors is adding more complexities to operation problem of electrical grids.

The advent of modern power electronics has brought tremendous impact on power systems [9]. Power electronic interfaces facilitate the peneteration of renewable energies into the smart grids [10]. With the help of distributed energy resources [11], advanced metering infrastructure, control [12] and communication technologies [13] and smart energy management [14], [15], reliability of the power systems can be improved which is a feature of smart grid. By employing DERs, operating paradigm of power system has been shifted from "load following" to "energy positioning" which enable system operating with no need of back up generating units run on idling[16]. The core uses of smart grids include advanced metering infrastructure, demand response or demand-side management, distributed generation sources and new energy, increasing the output power and gain, communication circuit system, energy storage, distribution automation, comprehensive awareness of the regional location, electric transportation [17]-[19]. Self-healing, which can prevent failures or restore interrupted loads as fast as possible, is one of the most important features of smart grid [20]. Self-healing in transmission systems is studied [21], [22] and recently, has been an attractive field of study for distribution system operators. In [23], decreasing the operation cost while the problem constraints are fulfilled is the aim of operators in this challenge. In [24]-[25], a comprehensive multi-level conceptual framework for the self-healing infrastructure is presented. In 2011, the problems associated with improving the accuracy of fault location were analyzed in order to move toward a self-heal network [26], [27]. In addition, [28] introduced a method for improving accuracy by weight adaption in robotics applications. Also, in [29] and [30], fault detection methods were introduced in power system networks demonstrating self-healing concept. In [31], size and location of DGs are optimized for loss minimization using an iterative-analytical method. In [32], simultaneous employment of V2G and wind power in scheduling and operation of power systems considering the uncertainty of wind generation is presented. In [33], [34], distribution automation is optimally implemented in order to make the distribution system self-heal. However, in none of the above studies self-healing feature of distribution grids is studied in the presence of renewable energy resources. This paper studies the self-healing characteristic of smart distribution grid from the viewpoint of both planning and operation considering the stochastic nature of some of the distributed energy resources. In the planning level, system is divided into different microgrids with the goal of loss minimization. In the operation level, self-healing control actions are decided based on the opted system configuration in the previous stage. The remaining of this paper is as follows: section II allocates to a conceptual framework for self-healing feature in smart distribution grids. Problem formulation is presented in section III. Problem methodology is discussed in section IV. Case study simulation is presented in section V, and finally, conclusions are drawn in section VI.

## II. SELF-HEALING CONCEPTUAL FRAMEWORK

When a fault occurs in the distribution grid, a self-heal system can divide the system into independent microgrids. The border of each microgrid from the other one can be distinguished by minimizing the active and reactive power imbalances in every single microgrid as shown in Fig. 1 [35].

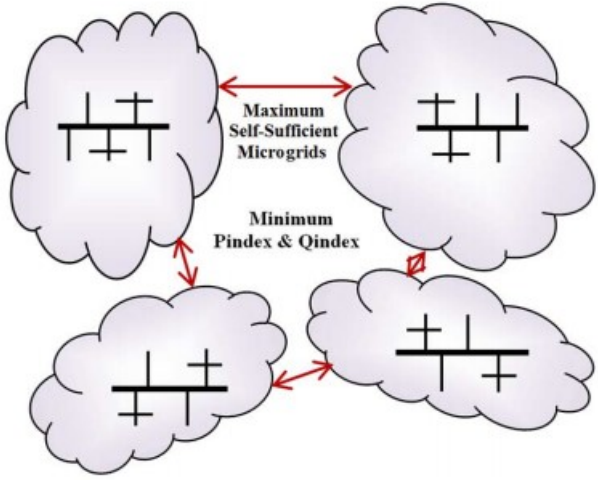

Fig. 1. Conceptual way of dividing power grid into various microgrids

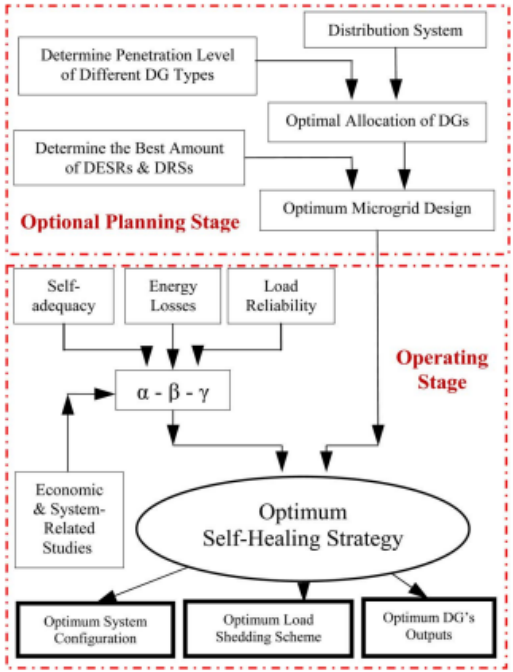

Fig. 2. Conceptual framework of the proposed approach

In other words, if minimizing power imbalances is done in the planning stage, power balancing between supply and demand will be more straightforward in the operation stage. In this paper, three different kinds of distributed energy resources including: wind turbine power plant, photovoltaic solar panels, and biomass generators are considered in the system. Then, optimally sizing and placement of this units are deployed in order to minimize total losses. When the optimal sizing and placement of distributed energy resources is implemented in the power grid, power imbalances in every single microgrid is minimized at the same time. This method of designing makes the smart distribution grid more reliable and robust against possible failures. The concept behind the proposed approach is depicted in Fig. 2.

When a fault occurs, self-healing feature of the system at the first stage prevents the fault from other parts of the system by isolating the faulted area. In the second stage, self-healing can restore as many customers as possible through system reconfiguration and utilizing the power of distributed energy resources.

## III. PROBLEM FORMULATION

The optimal operating points of the power system are computed based on the net-load (demands minus renewables) forecast. However, due to uncertain nature of renewables, scheduled operating points based on the forecasted net-load may be infeasible under actual net-load condition [36].

In order to have a robust system against contingencies, stochastic nature of wind turbines as well as solar photovoltaic panels should be taken into account. In order to accomplish this, wind speed and solar radiation; which are uncertain parameters, should be modeled when the optimization problem is solved using Weibull and Beta probability distribution functions, respectively [37], [38]. Three kinds of Distributed Generators (DGs) are considered in this paper which are biomass, solar, and wind power. The problem is solved for one day based on hourly time series data.

### A. Objective Function

The problem of optimal sizes and places of DGs is formulated in this section in order to have the minimum amount of annual losses. It should be noted that among all the buses, there are certain candidate buses based on the physical as well as environmental limitations. The objective function of this problem is as (1)

$$f = \sum_{n=1}^{n=N} P_{loss}^n \qquad (1)$$

where $P_{loss}^n$ is the total losses of being in the state $n$ and $N$ is the total states of the problem in a year. When $f$ is minimized, the system is ready to be divided into various separate microgrids.

### B. Stochastic Nature of Solar Power

Usually, the radiation of solar power in every hour of a day can be modeled using probability distribution function of Beta. The probability distribution function of Beta is a continuous function with two parameters of $\alpha \geq 0$ and $\beta \geq 0$ as shown in (2)

$$f_\beta(s) = \begin{cases} \dfrac{\Gamma(\alpha+\beta)}{\Gamma(\alpha)\Gamma(\beta)} \cdot s^{(\alpha-1)} \cdot (1-s)^{(\beta-1)} & 0 \leq s \leq 1 \\ 0 & Otherwise \end{cases} \quad (2)$$

where $s$ can be solar radiation parameter in $kW/m^2$. Average ($\mu$) and standard deviation ($\sigma$) of the Beta distribution function can be calculated using (3) and (4), respectively.

$$\beta = (1-\mu) \times \left(\frac{\mu \times (1+\mu)}{\sigma^2} - 1\right) \quad (3)$$

$$\alpha = \frac{\mu \times \beta}{1-\mu} \quad (4)$$

## C. Stochastic Nature of Wind Power

Wind speed is another stochastic parameter; which should be modeled, in order to have a more realistic solution for the aforementioned optimization problem. There are different ways of modeling the stochastic nature of wind speed; among which the Weibull probability distribution function is one the most common one. The utilized Weibull distribution function is as (5)

$$f_r(v) = \left(\frac{2v}{c^2}\right) \times e^{\left[-\left(\frac{v}{c}\right)^2\right]} \quad (5)$$

where $f_r(v)$ is the Weibull distribution function for the uncertain parameter of wind speed $(v)$ and $c$ is an index that is calculated as (6)

$$c \approx 1.128 \times v_m \quad (6)$$

where $v_m$ is the maximum wind speed based on the historical data in $m/s$.

## D. Biomass Power

The other considered DG in this paper is a small scale biomass. It has been assumed that the nominal power of biomass generator is fixed and is not dependent on the other stochastic parameters.

## E. Solar and Wind Power Outputs

The output power of a photovoltaic panel is highly dependent on the sun radiation, ambient temperature, and the features of the used module. Thus, the Beta distribution is generated for every time step and the power output is calculated using (7-11).

$$T_c^n = T_a + S_{avg}^n \left(\frac{T_{op} - 20}{0.8}\right) \quad (7)$$

$$I_c^n = S_{avg}^n (I_{sc} + K_i (T_c^n - 25)) \quad (8)$$

$$V_c^n = V_{oc} - K_v T_c^n \quad (9)$$

$$P^n(S_{avg}^n) = V_c^n I_c^n K_f N = K_f P_{max} \quad (10)$$

$$K_f = \frac{V_{mp} I_{mp}}{V_{oc} I_{sc}} \quad (11)$$

where $T_c^n$ is the cell temperature in state $n$ in °C, $T_a$ is ambient temperature in °C, $S_{avg}^n$ is the average of radiation in state $n$ in $w/m^2$, $T_{op}$ is the operational temperature in °C, $I_c^n$ is the cell current for state $n$ in $A$, $I_{sc}$ is short circuit current in $A$, $K_i$ is the current temperature coefficient in $A/°C$, $V_c^n$ is the cell voltage for state $n$ in $v$, $V_{oc}$ is the open circuit voltage in $v$, $K_v$ is the voltage temperature coefficient in $v/°C$, $P^n(S_{avg}^n)$ is the cell power for the average radiation of $S_{avg}^n$ in $W$, $K_f$ is the fulling coefficient of the solar panels, $N$ is total number of cells in the used photovoltaic array, and $V_{mp}$ and $I_{mp}$ are the voltage and current when the maximum power is produced, respectively.

On the other hand, the output power of a wind turbine is solely dependent on the wind speed and the turbine itself. Thus, Weibull distribution is generated for all time intervals and the output power can be calculated using (12)

$$P_w(v) = \begin{cases} 0 & 0 \le v \le v_{in} \\ P_n \frac{v - v_{in}}{v_n - v_{in}} & 0 \le v \le v_{in} \\ P_n & 0 \le v \le v_{in} \\ 0 & 0 \le v \le v_{in} \end{cases} \quad (12)$$

where $P_w(v)$ is the output power of the wind turbine at speed $v$ in $W$, $P_n$ is the nominal power of the wind turbine, $v_{in}$ and $v_{off}$ are the cut-in and cut-off speeds associated with wind turbine, respectively, and $v_n$ is the nominal speed of the wind turbine.

## F. Constraints of the Problem

If it has been assumed that the penetration percentage of solar power, wind power, and biomass is $S\%$, $W\%$, and $B\%$, respectively, certain constraints are imposed on the problem as (13-15)

$$P_{w,tot} = \frac{W}{100} \times P_{R,tot} \quad (13)$$

$$P_{s,tot} = \frac{S}{100} \times P_{R,tot} \quad (14)$$

$$P_{b,tot} = \frac{B}{100} \times P_{R,tot} \quad (15)$$

where $P_{w,tot}$, $P_{s,tot}$, and $P_{b,tot}$ are total power of the wind, solar, and biomass generators, respectively. Moreover, $P_{R,tot}$ is the total penetrated renewable power.

Furthermore, the bus voltages should be within a standard range as formulated in (16)

$$V_{min} \leq V \leq V_{max} \quad (16)$$

where $V_{min}$ and $V_{max}$ are the minimum and maximum allowable voltages for every single bus in the system, respectively. Also, $V$ is the final voltage of each bus.

## IV. SOLUTION APPROACH

The objective of this paper is to optimally allocate DGs and their sizes in order to minimize total losses in the system. There are also certain constraints imposed on the problem; which have been taken into consideration. In this paper, in order to find the voltages of each buses, there is a need to run a power flow. The utilized power flow of this paper is forward backward power flow; which is very popular for distribution systems. As formulated in the section III, the optimization problem of this paper is a non-linear problem which has been solved using Particle Swarm Optimization (PSO) method.

## V. CASE STUDY SIMULATION

### A. Case Study

In this paper, IEEE 33 bus test system is used as the case study in order to reveal the effectiveness of proposed method on the voltage profiles as well as on the total imposed costs. Moreover, as claimed before, the biomass generators are considered as dispatchable units. In this paper, nominal powers of the solar panel, wind turbine, and biomass generator are $200kW$, $400kW$, and $600kW$, respectively. Moreover, the penetration percentage of solar, wind, and biomass units are 5.73%, 11.46%, and 17.19%, respectively. Indeed, 34.38% renewable energy penetration is considered in aggregate. Also, as an exemplification, the wind velocity for a day is considered as Fig. 3. Solar radiation profile is also depicted in Fig. 4 for a specific day. Based on Fig. 3 and Fig. 4, the parameters of the Weibull and Beta distributions are estimated in order to have a more realistic stochastic model for wind and solar units.

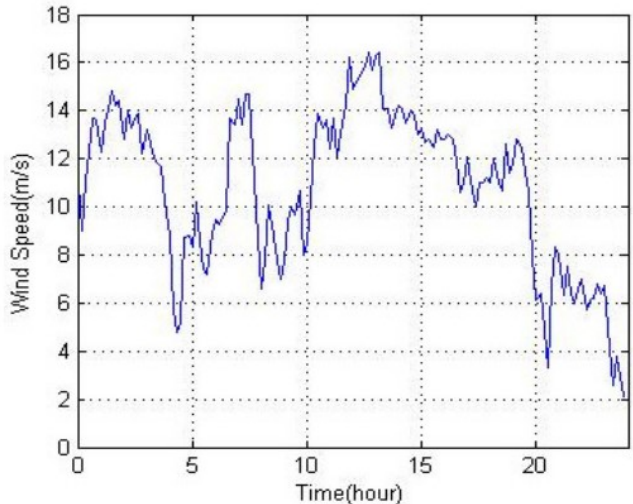

Fig.3. Wind speed profile in a specific day

Moreover, the other parameters of the wind and solar generators are brought in Tables I and II, respectively.

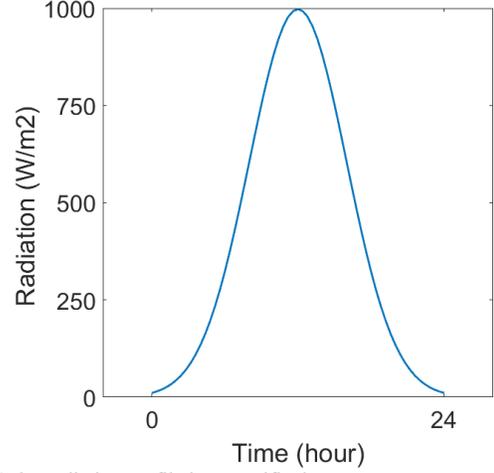

Fig.4. Solar radiation profile in a specific day

TABLE I. PARAMETERS OF WIND TURBINE

| Parameter | Value ($m/s$) |
|---|---|
| $v_{in}$ | 4 |
| $v_{off}$ | 25 |
| $v_n$ | 16 |

TABLE II. PARAMETERS OF SOLAR UNIT

| Parameter | Value |
|---|---|
| $K_i$ | $1.22 \ m \ A/°C$ |
| $K_v$ | $14.40 \ m \ v/°C$ |
| $I_{mp}$ | $4.76 \ A$ |
| $V_{mp}$ | $17.32 \ v$ |
| $I_{sc}$ | $5.32 \ A$ |
| $V_{oc}$ | $21.98 \ v$ |
| $T_c^n$ | $43 \ °C$ |
| $P_{max}$ | $75 \ W$ |

### B. Simulation Results

The problem of optimally allocation of DGs and their sizes with the goal of loss minimization is solved using PSO. After finding the optimal solution, the optimum places of DGs as well as their sizes are presented in Table III considering the stochastic nature of wind speed and solar radiation. It should also be noted that the total system losses has decreased from 317.6 to 127.75 kW. Moreover, the bus voltage profiles as depicted in Fig. 5, are improved up to a considerable degree.

TABLE III. Simulation results

| DG | Optimal places of DGs | Capacity (kW) |
|---|---|---|
| Wind | 8, 9, 14, 18, 28, 30, 31 | 75, 75, 75, 50, 25, 75, 25 |
| Solar | 4, 7, 17, 23, 26 | 50, 25, 50, 50, 25 |
| Biomass | 2, 3, 5, 11, 13, 15, 19, 25, 27, 29, 32, 33 | 50, 25, 50, 25, 25, 75, 25, 50, 100, 100, 25, 50 |

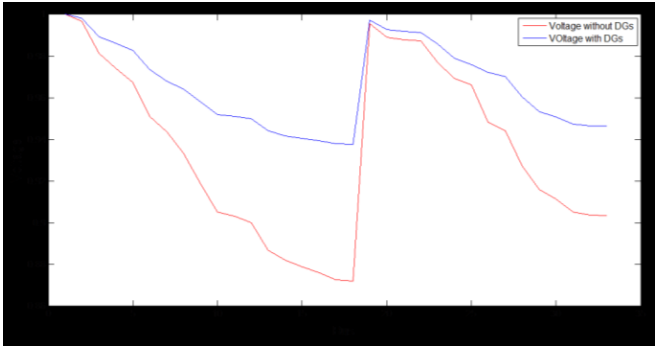

Fig.5. Voltage profiles

## VI. CONCLUSIONS

In this paper, optimally allocating and sizing of DGs are implemented with the goal of loss minimization in a microgrid. As a result, the reliability of the system has increased as a very pivotal characteristics of the microgrids. Moreover, it is shown that voltage profile of buses; which is very important especially in distribution grids, has been improved. It should be noted that the uncertain nature of renewable energy resources has been taken into consideration in order to have a more realistic model. Also, after adding the DGs into the system, system can be divided into various microgrids based on minimization of power imbalances between two different microgrids.